\documentstyle[11pt,aaspp4]{article}
\lefthead{SOFIA AND JENKINS}
\righthead{NEUTRAL ARGON IN THE ISM}
\begin{document}
\title{IMAPS Observations of Interstellar Neutral Argon and the
Implications for Partially Ionized Gas}
\author{Ulysses J. Sofia\altaffilmark{1}}
\affil{NRC, NASA/GSFC Code 681 Greenbelt, MD 20771}

\and

\author{Edward B. Jenkins}
\affil{Princeton University Observatory, Princeton, NJ 08544}

\altaffiltext{1}{Current address: Department of Astronomy and
Astrophysics, 
Villanova University, Villanova, PA 19085} 

\begin{abstract}
We use the absorption features from the resonance transitions of neutral
argon at 1048 and 1066\,\AA\ to determine interstellar argon abundances 
or their lower limits toward nine early-type stars. These features were
observed with the Interstellar Medium Absorption Profile Spectrograph
(IMAPS) along sight lines with low reddening and low fractional
abundances of molecular hydrogen. Four of the sight lines in our sample 
have reliably measured interstellar Ar abundances, of which three also
have good determinations of $N({\rm H~I})$.  We find that the
interstellar Ar is below its solar and B-star abundance with
respect to hydrogen toward $\zeta$ Pup, $\gamma^{2}$ Vel and $\beta$ 
Cen A with (logarithmic) reduction factors $D=-0.37 \pm 0.09\,{\rm
dex}$, $D=-0.18\pm 0.10\,{\rm dex}$, and $D=-0.61\pm 0.12\,{\rm dex}$,
respectively.

While Ar can condense onto the surfaces of dust grains in the interiors
of dense clouds, it is unlikely that argon atoms are depleted by this
process in the low-density lines of sight considered in this study. 
Instead, we propose that the relatively large photoionization cross
section of Ar makes it much easier to hide in its ionized form than H.
In regions that are about half ionized, this effect can lower Ar~I/H~I
by $-0.11$ to $-0.96\,{\rm dex}$, depending on the energy of the
photoionizing radiation and its intensity divided by the local electron
density.  We apply this interpretation for the condition of the gas in
front of $\beta$~Cen~A, which shows the largest deficiency of Ar.  Also,
we determine the expected magnitudes of the differential ionizations for
He, N, O, Ne and Ar in the partly ionized, warm gas in the local cloud
around our solar system.  For the local cloud and others that can be
probed by UV studies, the observed Ar~I to H~I ratio may be a good
discriminant between two possible alternatives, collisional ionization
or photoionization, for explaining the existence of partly ionized
regions.
\end{abstract}


\section{Introduction}\label{intro}

For diffuse, neutral gases in space, two fundamental processes influence
the pattern of abundances of different elements in free atomic form: (1)
the contributions from various nucleosynthetic sources and how the
outcomes are influenced by mixing and transport processes, and (2) the
removal of atoms as they are incorporated into dust grains.  There have
been numerous studies of the gas-phase abundances of elements in the
local region of our Galaxy \markcite{75, 328} (Cowie \& Songaila 1986;
Savage \& Sembach 1996), using information provided by UV interstellar
absorption lines in the spectra of stars.  These investigations have
generally focused on the second of the two processes \markcite{1063,
2722} (Jenkins, Savage, \& Spitzer 1986; Sofia, Cardelli, \& Savage
1994), using abundances in the Sun or nearby, early-type stars as a
comparison template for material that would be free of dust.  By
contrast, studies of very distant gas systems that produce absorption
lines in quasars have concentrated on the history of element production
\markcite{1162, 2665, 3087, 3446, 3447} (Pettini, Boksenberg, \&
Hunstead 1990; Pettini et al. 1994, 1997; Timmes, Lauroesch, \& Truran
1995; Lu et al. 1996), i.e., its character and rate, and have regarded
the depletions caused by dust formation as a possible complication that,
if present, must be compensated for.

It is generally recognized that the overall severity of a given
element's depletion is correlated with the stability of chemical bonds
of likely compounds in the grains.  This differential effect can be
interpreted to result from either the relative completeness of the
element's incorporation into grains in the outer envelopes of stellar
atmospheres \markcite{2878} (Field 1974), how rapidly the element
contributes to the growth of grains in dense interstellar clouds
\markcite{1254, 3390} (Snow 1975; Tielens \& Allamandola 1987), or the
relative resistance of different grain components due to destruction by
interstellar shocks \markcite{3391, 2725, 2783} (Seab 1987; Jones et al.
1994; Tielens et al. 1994).  Elements that have very strong bonds in
dust compounds, such as Al, Ti and Ca, are depleted by as much as 2 to 3
orders of magnitude \markcite{1388} (Jenkins 1987), while more volatile
elements, such as C, N and O, have depletions of only factors of 2 or 3
\markcite{14, 2858, 3346, 3438, 3466} (Hibbert, Dufton, \& Keenan 1985;
Meyer, et al. 1994; Cardelli et al. 1996; Meyer, Cardelli, \& Sofia
1997; Meyer, Jura, \& Cardelli 1997) or perhaps even less -- see below. 
Noble gases represent the greatest extreme of volatility, since they
generally do not form chemical bonds with solid compounds, and they can
adhere to grains only through the very weak van der Waals force
\markcite{3336} (Watson 1976).  This paper presents observations of the
interstellar abundances of one such noble gas, the element argon in its
neutral form.  One feature of this study is the fact that the lines of
sight have a low average gas density, and this further reduces the
likelihood that the abundances are modified by dust formation
\markcite{1388} (Jenkins 1987).

Neutral argon has two strong resonance lines in the far ultraviolet. 
Their f-values differ by a factor of 4 \markcite{2367} (Federman et al.
1992), which allows one to test for line saturation and apply corrective
measures (or realize that the saturation is hopelessly bad).  While the
interstellar Ar features are easily recognizable in most spectra, their
great strength turns out to be a liability because the lines become
strongly saturated when dense clouds are being investigated.  For
instance, for a velocity dispersion $b$ = $10\,{\rm km~s}^{-1}$ and
$N({\rm H})=10^{20}{\rm cm}^{-2}$, the weaker of the two Ar~I lines at
1066.66\,\AA\ would have a central optical depth $\tau_0=4$ if the argon
had a cosmic abundance ratio with respect to hydrogen.  The consequences
of such saturation are lessened if the line's velocity structure is
mostly resolved by the instrument, because one can interpret the
apparent optical depth as a function of velocity \markcite{110, 3184}
(Savage \& Sembach 1991; Jenkins 1996) instead of the much cruder
approach of just measuring the equivalent widths and using the standard
curve of growth to derive a total column density.  The observations
reported here take advantage of an instrument that can, in most
circumstances, resolve nearly all of the velocity detail of a feature,
which puts us in a better position to derive accurate column densities
than was possible with the {\it Copernicus\/} satellite \markcite{1393}
(Rogerson et al. 1973a).

The reference abundance of Ar derived for the Sun or B stars is now
understood with reasonable accuracy (see \S\ref{abundances}), and the
difference between the two is not very large.  The only major
uncertainty is whether or not the solar or B-star abundances are indeed
appropriate standards for comparison.  Recently accumulated evidence
\markcite{3346, 386, 3386, 3438, 3466} (Cardelli et al. 1996;
Fitzpatrick 1996; Meyer, Cardelli, \& Sofia 1997; Meyer, Jura, \&
Cardelli 1997; Sofia et al. 1997) indicates that some mildly depleted
elements do not follow the usual trend of depletions becoming stronger
in denser interstellar clouds \markcite{1063, 2773, 1388} (Spitzer 1985;
Jenkins, Savage, \& Spitzer 1986; Jenkins 1987), as we would expect if
these elements were major constituents in grains that are more abundant
and larger in dense regions because they are protected from interstellar
shocks and are allowed to grow rapidly \markcite{3390, 1380} (Tielens \&
Allamandola 1987; Jenkins 1989).  This phenomenon has raised the
possibility that the intrinsic abundances of heavier elements in the
local interstellar medium (in both solid and gaseous form) may be lower
than for the nearby stars, and that the depletions by grain formation
for some of them are relatively small in comparison to this
deficiency\footnote{In the current literature on interstellar
abundances, the word ``depletion'' usually implies the removal of atoms
by condensation onto grains.  In this article, we will likewise use this
word in the same restricted sense, and for more generalized
interpretations of atomic losses, without regard to the mechanism, we
will use ``deficiency.''  To emphasize the difference further in a
discussion of differential ionization effects in
\S\protect\ref{ionization}, we adopt the term $\Delta D_{\rm Ar}$ to
mean how much more of the neutral form of argon is reduced compared to
that of hydrogen, in contrast to a straight reduction $D$ that may occur
when the gases are in a fully neutral region but Ar condenses out onto
the grains.}.

A noteworthy property of argon is that, while its neutral form has an
ionization potential greater than that of hydrogen, its cross section
for photoionization above this threshold is extraordinarily high.  This
is of no consequence for an interstellar cloud that is thick enough to
have virtually all of its interior shielded from outside sources of
ionizing radiation by a large column of neutral hydrogen.  However the
situation for thinner clouds is different.  Since the absorption cross
section of hydrogen diminishes for energies well above the Lyman limit,
we naturally expect that relatively hard radiation must penetrate the
cloud and partially ionize the hydrogen and argon atoms.  In
circumstances where some or almost all of the hydrogen is ionized, the
fraction of argon that remains in neutral form can be considerably lower
than the corresponding fraction for hydrogen.  As a result, this element
may deceptively appear to be ``depleted'' relative to hydrogen if one
makes the simple-minded assumption that both elements are predominantly
neutral.  We will explore this issue in more detail in
\S\ref{ionization} and present an assessment of the conditions where
this effect is likely to be significant.

Our discussion starts with the observations and data reduction in
\S\ref{obs}, followed by the resulting Ar~I column densities
(\S\ref{col_dens}) and how they relate to measurements of neutral
hydrogen along the same lines of sight (\S\ref{abundances}).  In
\S\ref{previous} we briefly compare our results with those of previous
observations at much lower resolution.  On the theoretical side, in
\S\ref{depletion?} we will present in greater depth the arguments that
support our assertion that the apparent deficiencies of Ar~I relative to
H~I are unlikely to be caused by the depletion of Ar onto dust grains. 
We argue in \S\ref{ionization} that for our observing circumstances the
most likely reason for a reduction in Ar~I is because it is readily
photoionized by energetic radiation that is able to penetrate the
region.  We can use this effect to understand better the nature of the
partially ionized clouds.  Ar is a good tracer element to indicate how
strongly the apparent abundances of other elements could be influenced
by ionization, and this information is crucial for research on the
chemical evolution of the gases in different contexts.

\section{Observations \& Data Reduction}\label{obs}

The ultraviolet absorption line data used for
this study were obtained with the Interstellar Medium
Absorption Profile Spectrograph (IMAPS) during the first ORFEUS-SPAS
Space Shuttle mission in September 1993. IMAPS is a simple, 
objective-grating echelle spectrograph that can record the spectrum of a bright,
hot star in the wavelength interval 950 to 1150\,\AA, a spectral region
that is densely populated with atomic and molecular transitions from the
neutral and ionized interstellar medium (ISM) \markcite{1201, 1326, 75} (Morton
1975; Spitzer \& Jenkins 1975; Cowie \& Songaila 1986).  The
instrument was designed to achieve the high resolving power needed to
show most of the detailed velocity structures in the ISM. 
Jenkins, et al. \markcite{340} (1996) give a comprehensive
description of the IMAPS instrument, its flight on the first ORFEUS-SPAS
mission, and the methods of data correction and analysis. This paper 
includes observations in 9 sight lines for which the neutral argon resonance
transitions at 1048.220 and 1066.660\,\AA\ were recorded. These paths through 
the ISM are toward nearby, early-type stars with low reddening.
Table~\ref{targets} shows the list of target stars with their galactic
coordinates, spectral types, visual magnitudes, distances, B$-$V color
excesses, and line-of-sight fractions of molecular hydrogen.

\placetable{targets}
\begin{deluxetable}{lccccccc}
\tablecaption{Interstellar Sight Lines \label{targets}}
\tablewidth{0pt}
\tablehead{
\colhead{Star} & \colhead{l}   & \colhead{b}   &
\colhead{Type\tablenotemark{a}}
& \colhead{V\tablenotemark{a}} &
\colhead{Distance\tablenotemark{b}}  & \colhead{E(B$-$V)} & 
\colhead{Log f(H$_{2}$)\tablenotemark{c}}}
\startdata
$\gamma$ Cas     &123.58 &$-$2.15  & B0\,IVe       &2.47 &188  &0.13 
&$-$5.50\nl
$\epsilon$ Ori   &205.22 &$-$17.24 & B0\,Iae        &1.70 &412  &0.05 
&$-$3.91\nl
$\zeta$ Ori      &206.46 &$-$16.59 & O9.5\,Ibe      &2.05 &251  &0.05 
&$-$4.66\nl
$\kappa$ Ori     &214.52 &$-$18.50 & B0.5\,Iav      &2.06 &221  &0.04 
&$-$4.92\nl
$\alpha$ Eri     &290.84 &$-$58.79 & B3\,Vpe       &0.46 &44   &0.07    
& \ldots\tablenotemark{d}  \nl
$\zeta$ Pup      &255.98 &$-$4.71  & O5\,Iaf       &2.25 &429  &0.04
&$-$5.50  \nl
$\gamma^{2}$ Vel &262.81 &$-$7.70  & WC8+O7.5e   &1.78 &258  &0.05
&$-$5.51  \nl
$\beta$ Cru      &302.47 &3.18     & B0.5\,III      &1.25 
&108     &0.03 &\ldots\tablenotemark{d}  \nl
$\beta$ Cen A    &311.77 &1.25     & B1\,III       &0.61 &161   &0.02
&$-$6.44 \nl

\enddata

\tablenotetext{a}{From \protect\markcite{3005} (Hoffleit \& Jaschek
1982)}
\tablenotetext{b}{Based on the Hipparcos catalog parallaxes}
\tablenotetext{c}{f(H$_{2}$) = N(H$_{2}$) / [2N(H$_{2}$) + N(\ion {H}
{1})];
The \ion {H} {1} column densities are from Diplas \&
Savage \protect\markcite{2712} (1994), and the
H$_{2}$ column densities are from
Bohlin, Savage \& Drake \protect\markcite{1026} (1978).}
\tablenotetext{d}{No \ion {H}{1} or H$_{2}$ column densities are
measured toward this star.}
\end{deluxetable}

Once the images of the echelle spectra have been reduced and
appropriately coadded, the fluxes {\it vs.} wavelength must be
extracted. Our basic extraction method was discussed in
Jenkins et al. \markcite{340} (1996),
however we have altered some
of the procedures since the time of that article. Jenkins et al.
described a spectral extraction slit that had a uniform cross-dispersion
shape and spacing between orders. These parameters are important for the
extraction because the low intensity wings of adjacent echelle orders
overlap in the cross-dispersion direction. We used extraction slits with
predefined empirically determined cross-dispersion profile shapes which
varied slightly as a function of spectral order. We also allowed the
profile to be broadened or narrowed, and the distance between the orders
could vary slightly in order to better fit the data. These modifications
to the extraction slit were performed interactively during the
extraction.

\begin{figure}
\plotone{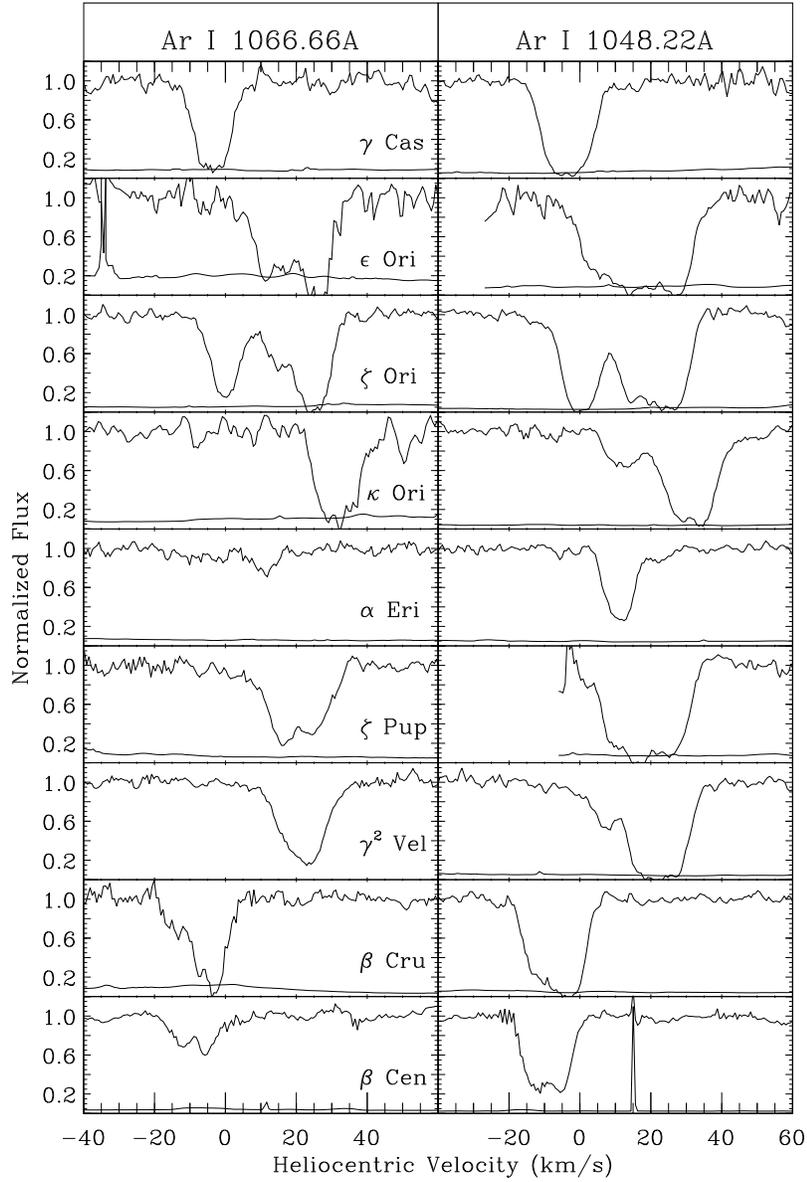}
\caption{Normalized flux versus velocity for the 
$\lambda\lambda$ 1048, 1066\,\AA\space transitions of neutral Ar in the 
interstellar medium toward our sample of stars. The lower curve in 
each frame represents the 2$\sigma$ statistical error in the data.
\label{intens_prof}}
\end{figure}

Our background level determination for the extraction procedure also
differed from that discussed in Jenkins et al.
\markcite{340} (1996).
Here we examined the regions which contain the least
contribution from the spectral orders, halfway between the order being
extracted and each of the adjacent orders. Since the shape of the
order's cross-dispersion profile is known, we are able to subtract the
contributions of the 2 adjacent orders from each of these interorder
regions. The remaining light was assumed to be the smooth background
which changes with position in the echelle dispersion direction. The
average of the two interorder points on either side of the order being
extracted was used as the background level and subtracted.  To test the
reliability of this method, we inspected the cores of lines that, from
their appearance, are likely to be fully saturated and hence have zero
intensity in their cores.  We found that the rms dispersion of these
zero levels was approximately 5\% of the local continuum flux level.
Figure~\ref{intens_prof} shows the extracted, normalized profiles of
\ion {Ar} {1} $\lambda\lambda$ 1048 and 1066\,\AA\space
toward the stars in our sample. 
The sampling interval for the curves in Figure~\ref{intens_prof}
is one half-pixel on the CCD since the data are put into an
oversampled format for reduction purposes
\markcite{340} (Jenkins et al. 1996).

\placefigure{intens_prof}

An interesting exercise is to compare the IMAPS absorption profiles with
the corresponding ones observed in the 1970's with {\it Copernicus\/}
\markcite{1393} (Rogerson et al. 1973a).  Although these two
instruments had
very different wavelength resolving powers ({\it Copernicus\/} could
resolve a velocity difference of about $15\,{\rm km~s}^{-1}$ compared 
to $4\,{\rm km~s}^{-1}$ for IMAPS), the
equivalent widths of the absorption lines should be the same if the
correct continua and background corrections were assigned in each case. 
Table~\ref{eqw} shows the
equivalent width measurements for the three \ion{N}{1} transitions near
$\lambda$ 1134\,\AA\space and the two \ion{Ar}{1} transitions at $\lambda$
1048\,\AA\space and $\lambda$ 1066\,\AA\space for $\gamma ^{2}$~Vel
\markcite{1205} (Morton \& Bhavsar 1979), 
$\zeta$~Pup \markcite{1203} (Morton 1978)
and $\epsilon$~Ori \markcite{1816} (Shull 1979). In the cases where the
absorption features reach near zero intensity (all of the \ion{N}{1} 
and \ion{Ar}{1} $\lambda$ 1048\,\AA\space features, as well as 
\ion {Ar} {1} $\lambda$ 1066\,\AA\space toward $\epsilon$ Ori), the
IMAPS equivalent widths agree well ($<$ 10\% difference) with those from
{\it Copernicus}. However the lines which do not approach zero 
intensity, i.e. the weaker line of \ion{Ar}{1} ($\lambda$
1066\,\AA ) toward $\zeta$ Pup and $\gamma^{2}$ Vel, show substantial
differences ($>$ 10\%) when comparisons are made between the IMAPS and
{\it Copernicus} measurements. Background levels were often difficult to
determine for {\it Copernicus} data, and we have no reason to believe
that the IMAPS background determinations for the \ion{Ar}{1} line at
1066\,\AA\ are worse than for any other region of the spectrum. We
therefore believe that our background determinations may be more
accurate for these weak lines than those determined for {\it Copernicus}
in this spectral region.

\placetable{eqw}
\begin{deluxetable}{llccccc}
\tablecaption{IMAPS vs {\it Copernicus} Equivalent Width Measurements 
\label{eqw}}
\tablewidth{0pt}
\tablehead{
\multicolumn{2} {c} { } &
\multicolumn{5} {c} {W$_{\lambda}$ (m\AA)} \\
\multicolumn {1}{l}{Star}&
\multicolumn {1} {l} {Instrument}&
\multicolumn {1}{c}{\ion {Ar} {1} $\lambda$ 1048.2}&
\multicolumn {1}{c}{\ion {Ar} {1} $\lambda$ 1066.6}&
\multicolumn {1}{c}{\ion {N} {1} $\lambda$ 1134.1}&
\multicolumn {1}{c}{\ion {N} {1} $\lambda$ 1134.4}&
\multicolumn {1}{c}{\ion {N} {1} $\lambda$ 1134.9} }

\startdata
$\epsilon$ Ori  & IMAPS& 112  & 76  & 128  & 137  & 141   \nl
      & {\it Copernicus\tablenotemark{a}}& 106  & 78  & 126  & 136  &
152   \nl
      & & & & & & \nl
$\zeta$ Pup     & IMAPS& 88   & 53 & 112  & 120  & 128   \nl
      & {\it Copernicus}\tablenotemark{b}&  89   & 65 & 114  & 123  &
131   \nl
            & & & & & & \nl
$\gamma^{2}$ Vel & IMAPS& 78   & 41  & 92  & 118  & 118   \nl
      & {\it Copernicus}\tablenotemark{c}& 72    & 54  & 99  & 113  &
107   \nl
\enddata
\tablenotetext{a}{\protect\markcite{1816} (Shull 1979).}
\tablenotetext{b}{\protect\markcite{1203} (Morton 1978).}
\tablenotetext{c}{\protect\markcite{1205} (Morton \& Bhavsar 1979).}
\end{deluxetable}

Jenkins \& Peimbert \markcite{362} (1996)
analyzed the telluric excited \ion{O}{1}
lines that appeared in the IMAPS spectrum of $\zeta$~Ori~A,
in order to determine the width of the
instrument's profile. Assuming that the lines are thermally broadened
for $T$ = 1000\,K, they found that the instrumental smearing function
for those observations was approximately a 3.5~pixel (FWHM) Gaussian, a
width that corresponds to a Doppler shift of 4\,km~s$^{-1}$.  It is
reasonable to assume that the resolving power for the other stars
observed during the ORFEUS-SPAS-I mission is approximately the same as
that for $\zeta$~Ori~A.  It is lower than what should
have been achievable in principle with IMAPS and the pointing stability
of the spacecraft.  We attribute the degradation to the relief of
mechanical stresses within a sticky bearing that supported the echelle
grating.  This problem led to small motions of the spectrum while the
exposures were underway.  The magnitude and character of this effect has
been discussed in detail by Jenkins, et al. \markcite{340}
 (1996).

\section{Column Densities of Argon}\label{col_dens}
\subsection{Method of Analysis}\label{method}

An important advantage of having observations that can resolve much of
the velocity structure is the ability to more explicitly detect and
possibly correct for line saturation.  In addition, sections of the
profile at different velocities may exhibit interesting behavior that would
otherwise be lost at low resolution.  While we could have used our
observations to define some combination of separate components and then
determined their amplitudes, velocity centroids and widths that are most
consistent with the data, we instead have chosen a more generalized
procedure.  We measured the intensity $I(v)$ in terms of the continuum
level $I_0$ and derived an apparent optical depth as a function of
velocity,
\begin{equation}\label{tau_a}
\tau_a(v) = \ln \Bigl( {I_0\over I(v)}\Bigr)~,
\end{equation}
a function whose interpretation has been described in some detail by
Savage \& Sembach \markcite{110} (1991) and Jenkins
\markcite{3184} (1996).
For the ideal cases where either the line
is very weak or we are sure that the instrument has resolved the finest
details in velocity, $\tau_a(v)$ gives an accurate depiction of a
differential column density per unit velocity through the relation
\begin{equation}\label{n_a}
N_a(v) = 3.768\times 10^{14}{\tau_a(v)\over f\lambda}~{\rm cm}^{-2}({\rm
km~s}^{-1})^{-1}~,
\end{equation}
where $f$ is the transition's $f$-value and $\lambda$ is expressed in
\AA.

In some circumstances, $N_a(v)$ derived through the use of
Eqs.~\ref{tau_a} and \ref{n_a} can give an underestimate for the true
column density per unit velocity $N(v)$.  If narrow, saturated
structures within the velocity profile are smoothed over by the
instrument, one will not fully recognize how closely the cores of the
features approached the zero intensity level before the smoothing took
place, thus leading to an incorrect representation of $N(v)$. 
Fortunately, we can sense when this is happening (and over what velocity
range) if we are able to observe two or more lines with different 
$f$-values and note that the weaker line gives a larger $N_a(v)$ than
corresponding values for the stronger line.  Jenkins \markcite{3184} (1996) has
described how one can use this information to derive a good
approximation to the true $N(v)$, aside from the fact that it has been
smoothed.  This method relies on a point-by-point correction technique
that mirrors the classical curve of growth analysis method that has been
used for the equivalent widths of lines.

A proper application of the $\tau_a(v)$ correction procedure requires
that the velocity scales of the strong and weak lines are accurately
matched.  In their IMAPS spectrum that showed many features of H$_2$ in
the spectrum of $\zeta$~Ori~A, Jenkins \& Peimbert \markcite{362}  (1996)
found that the velocities were internally consistent to within an rms
dispersion of only $0.5\,{\rm km~s}^{-1}$.  Since our method of
correcting for image distortions and deriving the wavelength scales is
identical to that of Jenkins \& Peimbert, we expect a similar accuracy
in this study.

Beyond the considerations discussed above, other key factors that
influence the accuracy of the column densities include a good knowledge
of the background and continuum levels in the region of the absorption
and, of course, accurate oscillator strengths for the transitions being
measured.  We have already discussed in \S\ref{obs} the probable
accuracy of our background subtractions. The stellar continua over the 
velocity breadth of the absorption features are quite smooth and well
determined for all of the stars. We found that a third-order 
polynomial fitted to the spectrum adjacent to the absorption
gave a satisfactory definition of the continuum. We 
estimate the continuum fitting error to be 0.35 times the rms deviation
of the data about the fitted continuum in the spectral region adjacent 
to the absorption line of interest [see Savage, Cardelli \& Sofia \markcite{2369}
(1992) 
for a justification of this error estimate]. For the two Ar transitions
we used the oscillator strengths $f_{1048}=0.257\pm 0.013$ and $f_{1066}=0.064\pm
0.003$ reported by Federman et al.
\markcite{2367} (1992).

\subsection{Results}\label{results}

\begin{figure}
\epsscale{.9}
\plotone{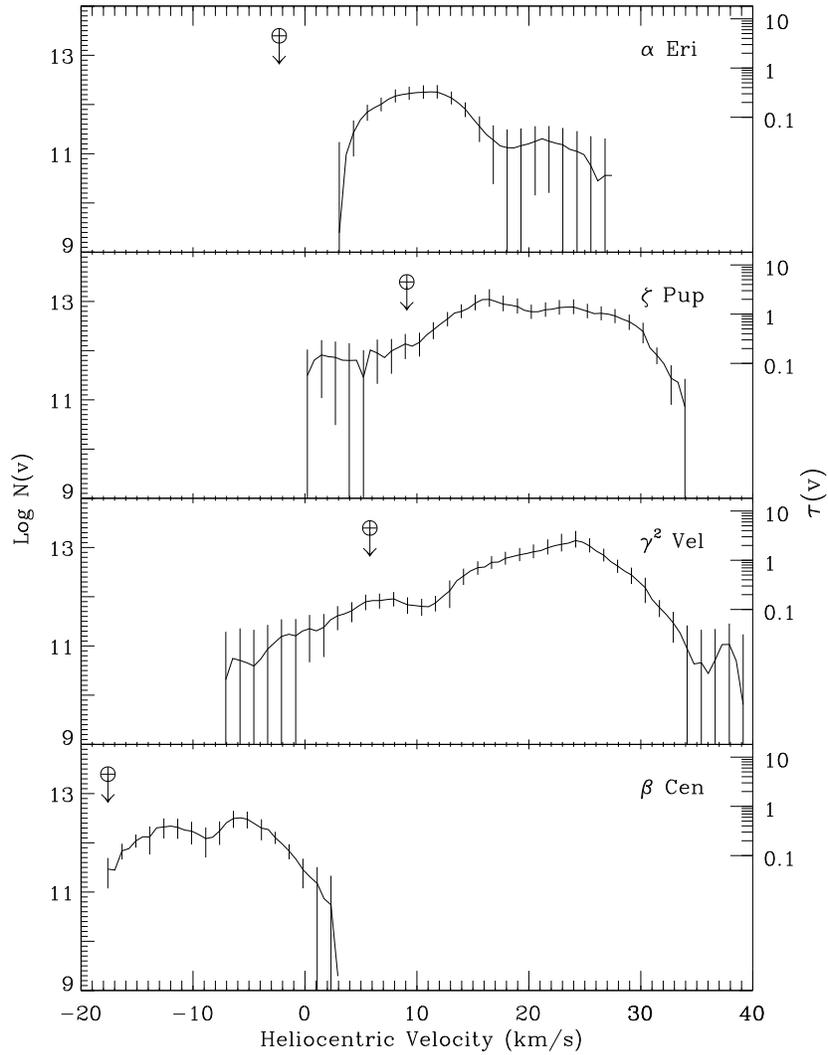}
\caption{Logarithmic column densities $N(v)$ (see
Eq.~\protect\ref{n_a}) as a function of 
velocity toward $\zeta$ pup, $\gamma^{2}$~Vel, $\beta$~Cen~A and $\alpha$~Eri. The
$\pm1 \sigma$ error bars 
include statistical, continuum fitting and background uncertainties. The 
optical depths on the right-hand side are representative of the $\lambda$
1066\,\AA\space 
\ion{Ar}{1} transition. The \earth\ symbol with an arrow in each frame indicates
the location 
of possible telluric Ar contamination. \label{na_prof}} 
\end{figure}

The optical depth correction procedure discussed in \S\ref{method} 
was applied to the Ar lines toward the stars in our sample. 
Valid corrections can not be made for line pairs in regions
where the weaker line has near zero intensity. Figure \ref{na_prof} 
shows the logarithmic column densities as a function of velocity for 
the four sight lines with corrected 
optical depth profiles which we believe do not contain any uncertain 
points ($\zeta$ Pup, $\gamma^{2}$ Vel, $\beta$ Cen A and $\alpha$ Eri); 
the remaining five sets of profiles could not be corrected across the 
entire velocity extent of the Ar absorption. Integrating Eq.~\ref{n_a} over
the velocity extent of the Ar absorption yields the column densities 
and limits listed in Table~\ref{abund}. The column densities were 
derived using the corrected optical depth profiles, and the limits are 
based on the uncorrected $\lambda$1066\,\AA\space optical depths. The
reported uncertainties are based on the statistical errors in the data,
the uncertainty in the oscillator strengths, the error in continuum
definition, and a background uncertainty of 5\% of the continuum flux level,
all of which were added in quadrature. No uncertainty relating to the
correction of the optical depth profiles is included. The differences
between the column densities measured after and before the optical depth
correction was applied to the weaker line are also shown in the third 
column of Table~\ref{abund} for those lines which could be corrected.
The differences between the pre- and post-correction column densities 
are quite small for these sight lines, all within the 1-$\sigma$ measurement
errors. This agreement gives us added confidence that our background
determinations are probably reliable.

\placefigure{na_prof}

\placetable{abund}
\begin{deluxetable}{lcccc}
\tablecaption{Integrated Argon Gas-Phase Abundances and Depletions
\label{abund}}
\tablewidth{0pt}
\tablehead{
\colhead{Star} &\colhead{Log N(Ar I)}     
&\colhead{Log N(Ar I) $-$ Log N$_{a}$(Ar I)\tablenotemark{a}}    & 
\colhead{Log[ N(H I) + 2N(H$_{2}$)]} &\colhead{D\tablenotemark{b}}}
\startdata
$\gamma$ Cas & $>$14.10 & \ldots & 
20.16$\pm$0.05\tablenotemark{c} &$>-$0.58\nl
$\epsilon$ Ori & $>$14.39 & \ldots & 
20.46$\pm$0.06\tablenotemark{c} &$>-$0.59\nl
$\zeta$ Ori    & $>$14.44 & \ldots & 
20.40$\pm$0.06\tablenotemark{c} &$>-$0.48\nl
$\kappa$ Ori & $>$14.18 & \ldots & 
20.54$\pm$0.04\tablenotemark{c} &$>-$0.88\nl
$\alpha$ Eri         &13.19 $\pm$ 0.09  & 0.00    
&$<$19.36\tablenotemark{d} & $>-$0.69\nl
$\zeta$ Pup          &14.14 $\pm$ 0.08  & 0.03    &19.99 $\pm$ 
0.02\tablenotemark{c} & 
$-$0.37 $\pm$ 0.09\nl
$\gamma^{2}$ Vel     &14.11 $\pm$ 0.08  & 0.08    &19.77 $\pm$ 
0.03\tablenotemark{c} & $-$0.18 $\pm$ 0.10\nl
$\beta$ Cru & $>$14.10 & \ldots & 
\ldots\tablenotemark{e} &\ldots\nl
$\beta$ Cen A        &13.45 $\pm$ 0.10  & 0.00    &19.54 $\pm$ 
0.06\tablenotemark{f} & $-$0.61 $\pm$ 0.12\nl

\enddata

\tablenotetext{a}{Numbers in this column indicate (the logarithm of) the
magnitude of the adjustment for unresolved, saturated features within
the absorption profile \protect\markcite{3184} (Jenkins 1996)-- see
Eq.~\protect\ref{n_a} and the paragraph that follows it.}
\tablenotetext{b}{As defined in Equation \ref{depl_eq}}
\tablenotetext{c}{The H column densities and errors are from a weighted
average of Bohlin, Savage \& Drake \protect\markcite{1026} (1978), and
Diplas \& Savage \protect\markcite{2712} (1994)}
\tablenotetext{d}{A limit from the measured interstellar S column 
density \protect\markcite{1365} (Rogerson et al. 1973b) as discussed in
the text}
\tablenotetext{e}{No estimate is available for the hydrogen column 
density}
\tablenotetext{f}{\protect\markcite{1760} (Rogerson \& York 1973)}

\end{deluxetable}

\section{Relative Abundances of Neutral Argon}\label{abundances}

In order to study the significance of our argon measurements, we must
compare them to measurements of other elements in the same gas.  When
possible, hydrogen is generally used as the comparison species.  The
cosmic abundance ratio of Ar to H is well determined.
Anders \& Grevesse \markcite{68} (1989)
give a logarithmic abundance for Ar/H of
$-$5.44 $\pm$ 0.10 which is based on corrected solar measurements, \ion
{H} {1} and \ion {H} {2} regions, and stars. Since our objective is to
study the abundances of Ar in the ISM, it would not be appropriate to
use a cosmic abundance which includes interstellar values.
Keenan et al. \markcite{1905} (1990) and
Holmgren et al. \markcite{3431} (1990)
find logarithmic ratios of $-$5.50 and $-$5.51 for
Ar/H in B star photospheres with uncertainties of $\pm$ 0.05 and $\pm$
0.10 respectively. Solar photospheric abundances determined from
correcting solar coronal observations have produced logarithmic ratios
of $-$5.46 $\pm$ 0.1 \markcite{68} (Anders \& Grevesse 1989)  and
$-$5.45 $\pm$ 0.08 \markcite{3432} (Meyer, 1989). We have taken the
weighted average of these B star and solar ratios to get a logarithmic
cosmic abundance ratio of $-$5.48 $\pm$ 0.04 for argon to hydrogen. 

The expected dominant ionization state for argon in most neutral hydrogen regions
is \ion{Ar}{1} since the ionization potential for neutral Ar is 15.76 eV
compared to 13.59 eV for H. Making the plausible initial assumption that all of
the argon associated with neutral H is \ion {Ar} {1} (an assumption that must
eventually be withdrawn, as discussed in \S\ref{ionization}), we determine the
reduction of Ar in the gas phase below its expected ratio to hydrogen through the
relation
\begin{equation}\label{depl_eq}
D = 
\log \left( \frac{N({\rm Ar~I})} {N({\rm H~I}) + 2N({\rm H_2})} \right)
_{observed} - 
\log \left( \frac{{\rm Ar}}{{\rm H}} \right) _{cosmic}.
\end{equation}
Table~\ref{targets} shows that the H$_{2}$ column densities toward the 
stars in our sample are very small compared to \ion{H}{1} and can therefore 
be neglected.
The interstellar neutral hydrogen column densities toward
$\gamma^{2}$~Vel and $\zeta$~Pup are $10^{19.77 \pm 0.03}$ and $10^{19.99 \pm
0.02}\,{\rm cm}^{-2}$, respectively, from a weighted average of the Bohlin,
Savage \& Drake \markcite{1026} (1978)
and Diplas \& Savage \markcite{2712} (1994) 
values. We adopt the determination $N({\rm H})=10^{19.54 \pm 0.06}\,{\rm cm}^{-2}$ 
toward $\beta$ Cen A from Rogerson \& York
\markcite{1760} (1973). A
reliable neutral hydrogen column density is not available toward $\alpha$ Eri
because the stellar H absorption badly contaminates the interstellar H 
spectral features. For this sight line we estimate a total H 
(i.e. 2H$_{2}$ + \ion{H}{1} + \ion{H}{2}) column density
from the \ion {S}{2} column density; with an ionization potential
of 23.33 eV, the \ion {S}{2} ion resides
in both the neutral and the ionized ISM toward $\alpha$ Eri. Assuming 
that S is undepleted \markcite{3329} (Fitzpatrick \& Spitzer 1997)
and that the cosmic S/H abundance is represented 
by B stars [Log(S/H) = $-$4.91 \markcite{3433} (Snow \& Witt 1996)], we 
find that the total logarithmic interstellar hydrogen column density 
toward $\alpha$ Eri is Log N(H) = 19.36 
for the S~II column 
density determined by Rogerson, et al.
\markcite{1365} (1973b). 
Based on the \ion{N}{1} and \ion{N}{2} abundances toward $\alpha$ Eri, 
it is possible that a substantial fraction of the H in the sight line is 
ionized.

For the three cases with good measurements of N(\ion{H}{1}) we 
determined the apparent logarithmic deficiencies $D$ of argon with
respect to hydrogen in the fifth column of 
Table \ref{abund}.
The accompanying errors include the uncertainty in the \ion {Ar}
{1} and \ion {H} {1} abundance determinations, and the uncertainty in
the cosmic Ar-to-H abundance ratio added in quadrature. The uncertain 
neutral H column density toward $\alpha$ Eri only allows a limit of
$D> -0.69$ toward this star. Lower limits for $D$ toward the 
other stars in our sample are given in Table \ref{abund}.

Due to the low orbit (295 km) of ASTRO-SPAS, there is a possibility
that telluric lines could contribute some additional absorption in our
spectra. Neutral argon is an
abundant species in the Earth's atmosphere, so we must
know the telluric absorptions' effects on the interstellar profiles. 
We calculated the column density of telluric neutral Ar that IMAPS would
observe through as a function of zenith angle using the global thermospheric model
of Hedin et al. \markcite{3434} (1974) with input
from the measured solar parameters for the
observation days (K. Schatten, private communication) and the assumption of
a spherical Earth. The expected Ar absorption
component in the atmosphere is generally small except for the largest
zenith angles, i.e., $N({\rm Ar~I})>10^{12}\,{\rm cm}^{-2}$ only when
$z>90\arcdeg$.  With our computed time-averaged column densities for
the target zenith angles of our exposures, we estimate that the values 
$< 10^{11}$, $2 \times 10^{12}$, $2 \times 10^{11}$ and $6 \times 
10^{10}$ atoms cm$^{-2}$ apply to the telluric contamination 
of the 1066\,\AA\space absorptions toward $\zeta$~Pup, $\gamma^{2}$~Vel,
$\beta$~Cen~A and $\alpha$~Eri, respectively.  These numbers should serve only as
an approximate indication of the possible telluric contamination, since the
predicted column densities could vary by
a factor of 2 in any given day.  The heliocentric velocity of the 
telluric absorption toward each of these stars is shown in
Figure~\ref{na_prof}. Any addition 
to Ar column densities from telluric lines would tend to make the Ar in a
sight line appear less deficient in the ISM than it actually is.

\section{Comparison with Previous Results}\label{previous}

The {\it Copernics} spectrometer was
used to observe the neutral interstellar argon lines 
toward several stars.  These earlier results have been summarized
by Meyer \markcite{3432} (1989).  Because of the
great strength of the absorption 
features (see \S\ref{intro}) and their tendency to saturate for moderately
dense lines of sight,
these measurements often led to large uncertainties in the 
Ar abundances.  For the very lightly reddened sight lines toward $\zeta$ Pup and 
$\gamma^{2}$ Vel, Morton \markcite{1203} (1978) and Morton \& Bhavsar
\markcite{1205} (1979) have reported
depletions of $-$0.31 $\pm$ 0.3 and $-$0.20 
$\pm$ 0.15, for the respective stars based on data from{\it Copernicus}.
These are very 
similar to the logarithmic deficiencies determined in this paper ($-$0.37 $\pm$
0.09 and 
$-$0.18 $\pm$ 0.10) even though the original $f$-values \markcite{1759} (Morton \&
Smith 1973)
and reference abundance \markcite{3435} (Withbroe 1971)
used for the {\it Copernicus} 
determinations have since been revised. It is the accuracy of the IMAPS 
data, however, that allows us to confidently conclude that the Ar 
is deficient with respect to our adopted reference abundance in our 
sample's sight lines, and that the relative reduction levels are not constant 
for this element in the ISM.

Federman et al. \markcite{2367} (1992) argue that the strength of the absorption
transitions 
requires that sight lines with small columns of neutral gas be used 
to determine reliable interstellar column densities of Ar. Using updated 
$f$-values and a new reference abundance, they reanalyzed the {\it Copernicus}
results from the low column density, low reddening sight lines toward 
$\alpha$ Vir \markcite{1783} (York \& Kinahan 1979) and $\lambda$ Sco
\markcite{1184} (York 1983)
and reaffirmed the previous conclusions that Ar/H is not below the solar value for 
the interstellar medium near the Sun. This seems to conflict with
our analysis of low column density, low reddening lines of sight which 
show that argon is indeed deficient in the neutral ISM in the Sun's 
vicinity. The fact that we find varied levels of Ar deficiencies, 
however, does not rule out such a difference as being valid.

\section {Discussion}\label{discussion}

\subsection{Can Ar Deplete onto Dust Grains?}\label{depletion?}

In studies of lightly depleted elements such as O, N and C
\markcite{3346, 3438, 3466} (Cardelli et al. 1996; Meyer, Cardelli, \&
Sofia 1997; Meyer, Jura, \& Cardelli 1997), the observed $D$ values
(about equal to $-$0.37, $-$0.09 and $-$0.39, respectively as compared
to solar values) are uniform over vastly different interstellar
conditions.  Such findings have led to the interpretations that the
abundances of these elements are not appreciably affected by the
formation of dust grains, at least over the range of cloud thicknesses
and internal densities surveyed.  Instead, it has been argued that the
total abundances of such elements in the local gas may be lower than
those of the Sun and possibly B stars \markcite{386} (Fitzpatrick 1996). 
While some of the deficiency of argon that we have detected may likewise
be explained partly by this phenomenon, the variation of $D$ shown for
the entries for $\zeta$~Pup, $\gamma^2$~Vel and $\beta$~Cen~A in
Table~\ref{abund} indicates that some additional process must be at
work.  We explore in this section the plausibility that argon could be
depleted by interacting with other materials in the medium.

In the interstellar medium, there are very few gas-phase compounds that
incorporate the argon atom, since this element is chemically inert. 
While Ar can form a stable hydride Ar\,H$^+$, Duley \markcite{3392}
(1980) argues that the concentrations of Ar\,H$^+$ in the interstellar
medium should be insignificant.\footnote{Duley \protect \markcite{3392}
(1980) assumed that Ar is entirely neutral and therefore the production
of Ar\,H$^+$ via the reaction ${\rm Ar}^+ + {\rm H}_2\rightarrow {\rm
Ar\,H}^+ + {\rm H}$ could be ignored.  In \S\protect\ref{ionization} we
challenge the notion that Ar is neutral in all H~I regions. 
Nevertheless, as indicated in Table~\protect\ref{targets}, there is
practically no H$_2$ present in our lines of sight.}  We are thus left
to consider only the direct attachment of single argon atoms onto the
surfaces of dust grains.

The characteristic e-folding time for the depletion of an element by
condensing onto dust grains is given by
\begin{equation}\label{t_d}
t_{depl.}=\big[\Sigma_d n({\rm H})\langle v\rangle S\big]^{-1}~.
\end{equation}
The quantity $\Sigma_d\approx 1\times 10^{-21}{\rm cm}^2$
\markcite{1854}  (Spitzer 1978, p. 162) is the cross sectional area of
the dust grains normalized to the local density of atomic hydrogen
$n({\rm H})$.  The average thermal speed of the atoms is $\langle
v\rangle$, and their sticking efficiency after impacting the grains is
$S$.  For the impact of a gas atom that is heavier than typical atoms on
the grain surfaces, an extrapolation of simple classical (soft cube)
models for the interactions indicate that $S$ should be of order unity
\markcite{3276} (Burke \& Hollenbach 1983).  We can reason that for
lines of sight with low average densities, such as those in this study,
significant depletions must have had to occur at some time in the past
when the material was in a dense cloud.  For instance, $t_{depl.}$ for
Ar at $T=100$\,K and $n({\rm H})=10^4\,{\rm cm}^{-3}$ is about
$1.4\times 10^5$\,yr, a value that is considerably shorter than a usual
cloud lifetime of $\approx 10^7$\,yr against collisions with other
clouds \markcite{1854} (Spitzer 1978, p. 231) or the average interval of
$\approx 10^8$\,yr between the passages of shocks that can destroy the
grains \markcite{3441, 2783} (Seab \& Shull 1986; Jones et al. 1994).

We note in passing that neutral argon has an ionization potential of
15.76 eV, i.e., a value greater than that of hydrogen.  As a
consequence, practically all of the argon atoms are neutral within H I
regions that have $n({\rm H})\gtrsim 10^4\,{\rm cm}^{-3}$ because these
clouds are generally thick enough to have their internal portions
shielded from external sources of ionizing radiation (this does {\it
not} apply to regions of lower density however, see \S\ref{ionization}). 
Because the atoms are neutral, the electrical charge of the grains is
not a complicating factor that must be entered in calculations of an
accretion rate for argon.  Also, the interaction of Ar as it comes in
contact with a grain is not made more complicated by the energy
liberated by the recombination of an ion or the creation of a chemical
bond.

For accretion to work, it must not be overwhelmed by processes that tend
to remove the atoms from the dust grain surfaces.  The Ar atoms on such
surfaces are bound by only a physical attachment caused by the van der
Waals interaction.  Owing to the weakness of this binding, these atoms
are susceptible to evaporation.  The residence of an atom on a surface
can be terminated by the rare thermal fluctuations in the lattice at the
dust temperature $T_d$ that can overcome the binding energy $Q$ at the
surface of the grain.  The mean time interval for this escape to occur
is given by
\begin{equation}\label{t_ev}
t_{ev}=\nu^{-1}\exp\left({Q\over kT_d}\right)
\end{equation}
where $\nu\approx 10^{12}{\rm s}^{-1}$ is the vibration frequency of the
grain lattice and $Q/k$ is estimated to be about 900\,K for Ar
\markcite{3336} (Watson 1976).  Draine \markcite{3452} (1994) has
estimated a characteristic grain temperature $T_d=18$\,K in the diffuse
interstellar medium from the infrared emission spectrum observed by
Wright et al. \markcite{3453} (1991) [see also Hauser, et al.
\markcite{3454} (1984)].  Substituting the values above for $\nu$,
$Q/k$, and $T_d$ into Eq.~\ref{t_ev} leads to the large inequality
$t_{ev}=160\,{\rm yr}\ll t_{depl}$.  For very small grains, the
situation is even worse: the discrete nature of the photoabsorptions
that heat the grains leads to significant, transient excursions in $T_d$
in the positive direction.  For instance, graphite or silicate grains
with diameters of order 50\,\AA\ or less spend a nonnegligible fraction
of the time at temperatures $T_d>30$\,K \markcite{3393} (Guhathakurta \&
Draine 1989), where $t_{ev}<10\,{\rm s}$!  Larger grains, however,
should not have such large temperature fluctuations, and their
equilibrium temperatures of around 15\,K [for grains with a radius of
around $0.5\mu$m \markcite{1291} (Draine \& Lee 1984)] lead to
$t_{ev}=4\times 10^6\,{\rm yr}$ which is much larger than the value
$t_{depl}=1.4\times 10^5\,{\rm yr}$ given earlier.

Another means for returning Ar to the gas phase is photoejection.  The
atoms on a grain surface can absorb the ambient stellar radiation at
wavelengths very near the 1048 and 1066\,\AA\ resonance lines.  For the
excited atoms a new interaction potential is created that ultimately
results in a breaking of the bond with the grain, either because it is
repulsive or causes an increase in the separation that can not be
accommodated by the ground-state potential when the atom eventually
decays \markcite{3394} (Watson \& Salpeter 1972).  For an average
interstellar radiation flux $4\pi\lambda J_\lambda=2\times 10^{-3}\,{\rm
erg~cm}^{-2}{\rm s}^{-1}$ at wavelengths near the Ar transitions
\markcite{3271} (Mathis, Mezger, \& Panagia 1983), we calculate that the
photon excitation (and ejection) rate $\beta=3\times 10^{-10}\,{\rm
s}^{-1}=(1/100\,{\rm yr})$.

The effect of this rapid ejection process can be overcome if, during the
time they were being accreted, the Ar atoms were smothered by molecules
or other atoms in a time scale that was short compared with
$\beta^{-1}$.  For instance, oxygen atoms (250 times more abundant than
argon) or oxygen-bearing compounds (principally water ice) could cover
argon atoms after their initial attachment to the grains.  Laboratory
experiments have demonstrated that significant quantities of Ar can be
trapped during the accumulation of solid H$_2$O on a cold surface
\markcite{3410} (Bar-Nun et al. 1987).  The time $t_c$ needed to cover
an Ar atom should be of order $(3\times 10^{14}\,{\rm s})/n({\rm H})$
for $T=100$\,K \markcite{3394} (Watson \& Salpeter 1972).  To retain
about one-half of the Ar striking the grains, it should be sufficient,
for example, to have a cloud whose interior has a density $n({\rm
H})=10^4\,{\rm cm}^{-3}$ that is shielded by about 0.5 magnitudes of
visual extinction (reducing $\beta$ to a value of about $3\times
10^{-11}\,{\rm s}^{-1}$).

In principle, we could consider the possibility that argon atoms are
locked inside coatings of water ice that accumulated at some previous
time when the grains were inside a very dense cloud.  If this were true,
the argon would not be quickly liberated by photodesorption after the
cloud had dissipated.  However, observations of the $3.05\mu$m ice-band
absorption feature indicate that ice-coated grains exist only within the
densest portions of compact clouds.  For such clouds, the approximate
linear trends in $\tau$(3.05) vs. A$_{\rm V}$ extrapolate to zero ice
absorption at A$_{\rm V}=2.6-5$ in different surveys \markcite{3425,
3427, 3426} (Whittet et al. 1988; Eiroa \& Hodapp 1989; Smith, Sellgren,
\& Brooke 1993).  A star with very strong foreground dust absorption
over a distance of 2.1\,kpc (VI~Cygni~No.12 with A$_{\rm V}=9.7$) shows
no detectable ice band absorption \markcite{3428} (Gillett et al. 1975). 
It therefore appears unlikely that our lines of sight hold grains that
are coated with ice, and thus Ar probably has little chance of being
bound to the grains.

From the above considerations, we conclude that Ar should be able to
experience some depletion well inside clouds that are optically thick to
radiation at the wavelength of the Ar transitions (the same ones that we
can observe), and that have large internal hydrogen densities and grains
with large diameters.  However, it appears likely that once the material
is returned to the low density phase of the ISM where we can observe the
absorption lines, the argon atoms are rapidly returned to the free
atomic form.

\subsection{The Ionization of Ar in Thin H~I Clouds}\label{ionization}

There is now good evidence for the existence of clouds that contain
significant amounts of H~I but with internal electron densities in the
general range $0.01 \lesssim n(e)\lesssim 0.1\,{\rm cm}^{-3}$ and
temperatures $10^3 \lesssim T \lesssim 10^4$\,K \markcite{2462, 250,
2701, 3329} (Spitzer \& Fitzpatrick 1993, 1995; Fitzpatrick \& Spitzer
1994, 1997).  For these high-temperature clouds it is clear that most of
the electrons must come from the ionization of hydrogen (rather than
only from those elements with an ionization potential less than that of
hydrogen).  Sciama \markcite{3443} (1997) has presented some arguments
that support the idea that the observations are indeed registering
partially ionized regions, rather than fully ionized ones that are
kinematically associated with neutral gases.  The presence of clouds
that have partially ionized material seems to be plausible on
theoretical grounds \markcite{3413} (Domg\"orgen \& Mathis 1994),
although the pervasiveness of such gas is constrained by an observed
upper limit for the ratio of the emission lines [O~I]
$\lambda$\,6300/H$\alpha$ \markcite{3423} (Reynolds 1989).  Probable
external sources of ionizing photons for clouds with a hydrogen column
density $N({\rm H})\approx 10^{18}{\rm cm}^{-2}$ include recombination
radiation from nearby, more fully ionized regions and the occasional
early B-type stars that are unobscured by other clouds: a familiar
example of the latter is the action of ionizing photons from
$\epsilon$~CMa that can penetrate and partially ionize the neutral cloud
around the Sun \markcite{298} (Vallerga \& Welsh 1995) -- see
\S\ref{lism}.  For thicker clouds, only the more energetic radiation
from white dwarf stars \markcite{1492, 3171} (Dupree \& Raymond 1983;
Dupuis et al. 1995), the hottest O-type stars, or emission from the
surrounding hot gas \markcite{1703, 1834, 3387} (Bloch et al. 1986;
Cheng \& Bruhweiler 1990; Snowden et al. 1997) or an evaporation
interface \markcite{1773} (Slavin 1989) can overcome the opacity of the
outer neutral material well enough to cause some ionization of hydrogen
and helium in the interior.  We now examine what the relative ionization
of argon should be in such regions, on the premise that it can markedly
influence the ratio of observed H~I to Ar~I.  [This phenomenon has also
been investigated for elements that can be depleted onto dust by
Cardelli, Sembach \& Savage \markcite{2886} (1995), Sembach \& Savage
\markcite{2960} (1996) and Welty, et al. \markcite{3437} (1997).]

The ionization potential of neutral argon at 15.76\,eV is slightly
higher than that of hydrogen, which means that argon, like hydrogen,
will have appreciable photoionization only for clouds that are thin
enough to admit some photons from the outside.  The recombination
coefficient of argon is about the same as that of hydrogen, but its
cross section for photoionization is about one order of magnitude higher
at most energies (see Fig.~\ref{Pfig}: the quantity $P$ is defined in
Eq.~\ref{Peq}).  In our development of the equation that expresses the
relative deficiencies of neutral hydrogen and argon caused by
ionization, we will begin with the simplest case where all of the
electrons come from the ionization of hydrogen and there are no charge
exchange reactions.  Once the basic equation has been outlined, we will
move on to consider complications that arise from charge exchange,
doubly-charged ions, and the ionization of helium.

\subsubsection{Simplest Case}\label{simple}

\begin{figure}
\epsscale{1.1}
\plotone{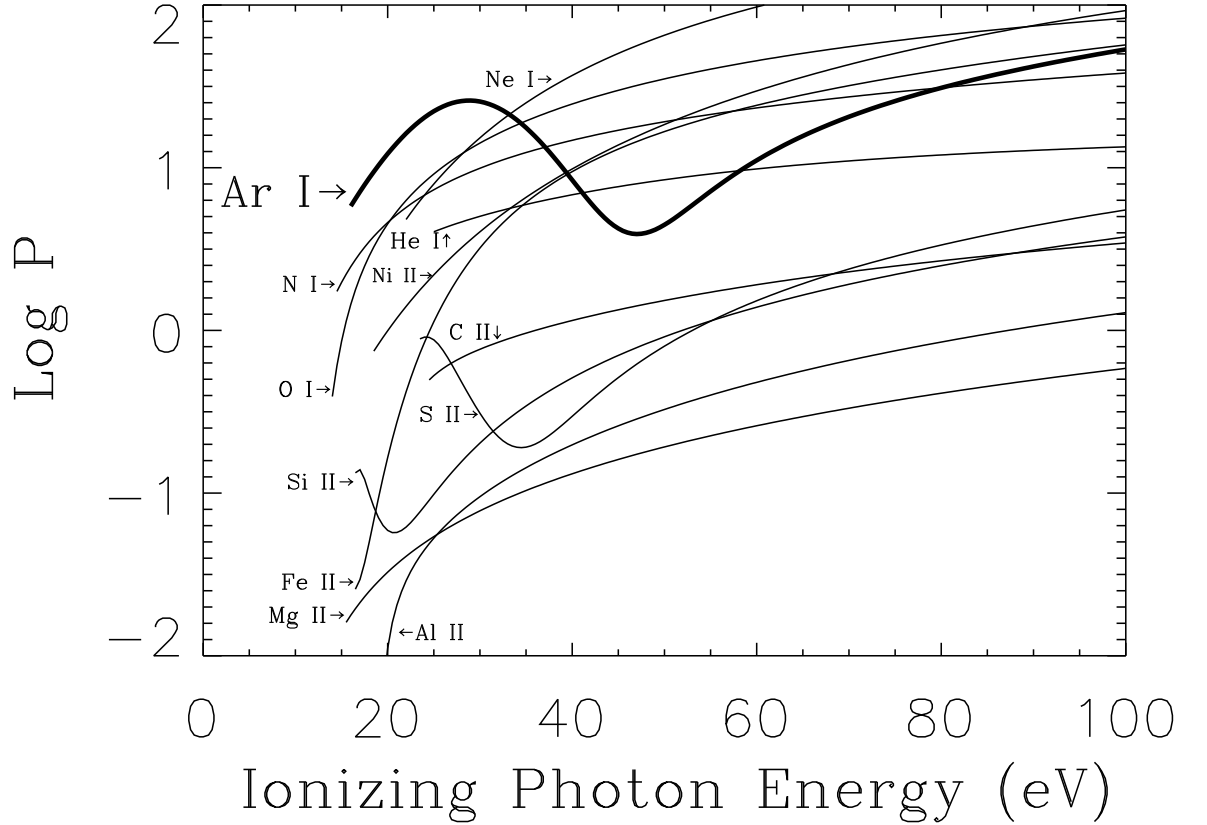}
\caption{Behavior of $\log P$ (as defined for Ar in Eq.~\protect\ref{Peq}) as a
function of energy of the photoionizing radiation, for a partially ionized gas at
temperature $T=5000$\,K.  The trend for argon, $P_{\rm Ar}$, is shown by a heavy
solid line.  $P$ values for other elements commonly observed in the ISM (plus He)
are shown by lighter lines.  The use of $P$ in Eq.~\protect\ref{DeltaD} will give
an approximate answer for $\Delta D$, except for O and N which are heavily
influenced by charge exchange reactions with H.  More accurate values may be
obtained by solving the equations in \S\protect\ref{complex} and replacing $P$
with $P^\prime$.  Representative values of $P^\prime$ for argon for various
physical conditions are given in Table~\protect\ref{Pprimes}.\label{Pfig}}
\end{figure}

Photoionization reduces the amount of neutral hydrogen by a factor
\begin{equation}\label{delta_H}
\delta_{\rm H}={n({\rm H})\over n({\rm H})+n(e)}=\left( 1+{n(e)\over
n({\rm H})}\right)^{-1}
\end{equation}
where in this instance $n({\rm H}^+)=n(e)$.  For a hydrogen ionization
rate $\Gamma ({\rm H})$ and recombination rate\footnote{In giving
numerical results, we use the total recombination rate to all levels
except $n=1$, on the assumption that the region is optically thick to
the recombination radiation at energies just above the Lyman limit.  For
very thin regions, it would be more appropriate to include the
recombination coefficient to all levels (larger by 50\% if
$10^3<T<10^4$\,K), since a Lyman limit photon will escape the region
instead of re-ionizing another hydrogen atom.} $\alpha ({\rm H})$ the
electron density is given by
\begin{equation}\label{n(e)1}
n(e)=\left({\Gamma({\rm H})n({\rm H})\over \alpha({\rm
H})}\right)^{\onehalf} ~.
\end{equation}
The recombination coefficient and ionization rate that apply to argon
determine the outcome for the reduction in neutral argon,
\begin{equation}\label{delta_Ar1}
\delta_{\rm Ar}={n({\rm Ar})\over n({\rm Ar})+n({\rm Ar}^+)}=\left(
1+{\Gamma({\rm Ar})\over n(e)\alpha({\rm Ar})}\right)^{-1}~.
\end{equation}
Substituting Eq.~\ref{n(e)1} into Eq.~\ref{delta_Ar1} gives
\begin{equation}\label{delta_Ar2}
\delta_{\rm Ar}=\left[ 1+{\Gamma({\rm Ar})\over \alpha({\rm
Ar})}\left({\alpha({\rm H})\over \Gamma({\rm H})n({\rm
H})}\right)^{\onehalf}\right]^{-1}=\left[ 1+\left({\Gamma({\rm
Ar})\alpha({\rm H})\over \Gamma({\rm H})\alpha({\rm
Ar})}\right){n(e)\over n({\rm H})}\right]^{-1}~.
\end{equation}
We define a quantity $P_{\rm Ar}$ such that
\begin{equation}\label{Peq}
P_{\rm Ar}\equiv \left({\Gamma({\rm Ar})\alpha({\rm H})\over \Gamma({\rm
H})\alpha({\rm Ar})}\right)
\end{equation}
so that the logarithmic difference in deficiencies is given by the
simple expression
\begin{equation}\label{DeltaD}
\Delta D_{\rm Ar}=\log \delta_{\rm Ar} - \log \delta_{\rm
H}=\log\left({1+{n(e)\over n({\rm H})}\over 1+P_{\rm Ar} {n(e)\over
n({\rm H})}}\right)~.
\end{equation}
To give a general impression of the expected values of $P$ under
different circumstances for argon and other commonly observed elements,
we show in Fig.~\ref{Pfig} plots of $\log P$ as a function of the energy
of the photoionizing radiation, assuming it is monoenergetic.  For
elements that are expected to be mostly singly ionized in H~I regions,
the $P$ values describe the magnitudes of the shift from singly to
doubly ionized forms.  We calculated the photoionization cross sections
from the analytic approximations of Verner, et al. \markcite{3404}
(1996) [and Verner \& Yakovlev \markcite{3408} (1995) for Ni].  The
recombination coefficients were evaluated from the parameters for the
fitting equations given by Shull \& Van Steenberg \markcite{1936} (1982)
[and Aldrovandi \& P\'equignot \markcite{1934} (1974) for Al].  In real
circumstances for the ISM, one must calculate explicitly the
photoionization rates by integrating over energy the flux times the
cross sections.  In deriving values of $P$ for Fig.~\ref{Pfig} we
assumed that $T=5000$\,K.  The recombination coefficients for different
elements change with temperature in a very similar way (except for
temperatures where dielectronic recombination becomes important), so $P$
will not change appreciably for temperatures that are somewhat different
than 5000\,K.

\placefigure{Pfig}

\subsubsection{More Complex Circumstances}\label{complex}

Eq.~\ref{DeltaD} is an oversimplication for energies above the
ionization potential of He, since an appreciable fraction of the
electrons can come from the ionization of this element. [Note that
Fig.~\ref{Pfig} shows that $P_{\rm He}$ is about 10, and this number
times the cosmic abundance ratio (He/H)~=~0.1 is 1].  Also, the earlier
development neglected the influence of charge exchange reactions and
photoionization to even higher stages.  In this section, we will
evaluate more realistic ionization ratios that take these considerations
into account, but we will express them by retaining the simple form of
Eq.~\ref{DeltaD} and replacing the simple values of $P_{\rm Ar}$ from
Eq.~\ref{Peq} with more correct ones (to be denoted $P_{\rm
Ar}^\prime$).  This representation is particularly useful since $n(e)$
and $n$(H) can often be determined by independent means, such as by
observing the relative populations of atoms and ions in excited
fine-structure levels \markcite{1847} (Bahcall \& Wolf 1968).

The ionization balance equations that yield the densities $n(X)$,
$n(X^+)$ and $n(X^{++})$ of the three lowest ionization stages of an
element $X$ (either He or Ar) in the presence of hydrogen,
\begin{equation}\label{equi1}
\big[\Gamma(X)+C(X^+)\big]n(X)=\big[\alpha(X)n(e)+C^\prime(X^+)n({\rm
H})\big]n(X^+)
\end{equation}
\begin{equation}\label{equi2}
\Gamma(X^+)n(X^+)=\big[\alpha(X^+)n(e)+C^\prime(X^{++})n({\rm
H})\big]n(X^{++})
\end{equation}
may be combined to yield the ionization fractions
\begin{eqnarray}\label{n0}
f_0(X) & \equiv & {n(X)\over n(X)+n(X^+)+n(X^{++})}\nonumber\\
&=&\left( 1\, +\, {\big[\Gamma(X)+C(X^+)\big]
\Big\{\Gamma(X^+)+\big[\alpha(X^+)n(e)+C^\prime(X^{++})n({\rm
H})\big]\Big\}\over \big[\alpha(X)n(e) + C^\prime(X^+)n({\rm
H})\big]\big[\alpha(X^+)n(e)+C^\prime(X^{++}) n({\rm
H})\big]}\right)^{-1}~,
\end{eqnarray}
\begin{eqnarray}\label{n++}
f_{++}(X) & \equiv & {n(X^{++})\over n(X)+n(X^+)+n(X^{++})}\nonumber\\
&=&{1-f_0(X)\over 1+\big[\alpha(X^+)n(e)+C^\prime(X^{++})n({\rm
H})\big]/\Gamma(X^+)}~,
\end{eqnarray}
and
\begin{equation}\label{n+}
f_+(X)\equiv {n(X^+)\over n(X)+n(X^+)+n(X^{++})}=1-f_0(X)-f_{++}(X)~,
\end{equation}
where the charge exchange rate constants $C^\prime(X^+)$ and
$C^\prime(X^{++})$ refer to the reactions $X^++{\rm H}\rightarrow X+{\rm
H}^+$ and $X^{++}+{\rm H}\rightarrow X^++{\rm H}^+$, respectively. [For
$T<10^4$\,K the rate constants $C(X^+)$ and $C(X^{++})$ for the reverse
reactions are very small for He and Ar since their ionization potentials
are greater than that of hydrogen by more than 1\,eV.  Nevertheless we
leave the $C(X^+)$ term in Eqs.~\ref{equi1} and \ref{n0} because it is
very important for O and of some significance for N, both of which will
be analyzed in \S\ref{lism}.]  For regions that are partially ionized,
the populations of atoms in high stages of ionization are severely
limited by their large charge exchange reaction rates with H (an effect
that makes it safe for us to disregard these higher stages in the
equilibrium equations).  Eqs.~\ref{n0}$-$\ref{n+} with $X$=He must be
solved simultaneously with the ionization balance for hydrogen,
\begin{mathletters}
\begin{equation}
{n({\rm H}^+)\over n({\rm H})}={\Gamma({\rm H})+Y\over \alpha({\rm
H})n(e)-Y}
\end{equation}
with
\begin{equation}
Y=0.1n({\rm H})\big[C^\prime({\rm He}^+)f_+({\rm He})+C^\prime({\rm
He}^{++})f_{++}({\rm He})\big]~,
\end{equation}
\end{mathletters}
together with the constraints
\begin{equation}\label{He_total}
n({\rm He})+n({\rm He}^+)+n({\rm He}^{++})=0.1n({\rm H})\left[1+{n({\rm
H}^+)\over n({\rm H})}\right]
\end{equation}
and
\begin{equation}\label{n(e)}
n(e)=n({\rm H}^+)+n({\rm He}^+)+2n({\rm He}^{++})~.
\end{equation}

A satisfactory method for obtaining simultaneous solutions to
Eqs.~\ref{n0}$-$\ref{n(e)} is to start with $\Gamma({\rm He})$ and
$\Gamma({\rm He}^+)$ equal to zero, and then iterate on the answers with 
gradual increases in the two helium ionization rates up to the final,
correct values.  After obtaining the final results for the coupled
hydrogen and helium ionization balances, one can solve for the argon
ionization balance with Eqs.~\ref{n0}$-$\ref{n+}.  These results can in
turn be used to derive the more accurate versions of $P_{\rm Ar}$,
\begin{equation}\label{Pprime}
P_{\rm Ar}^\prime=\left\{{\left[1+{n(e)\over n({\rm H})}\right]n({\rm
H})\over f_0({\rm Ar})[n({\rm H})+n({\rm H}^+)]}-1
\right\}\left[{n(e)\over n({\rm H})}\right]^{-1}~.
\end{equation}
Table~\ref{Pprimes} lists our calculations of $P_{\rm Ar}^\prime$ (and
$P_{\rm Ar}$ for comparison) for monoenergetic ionizing fields at
various energies.  In many cases the energies were chosen to be just
below or above an ionization edge for He or Ar.  Sources for the
photoionization cross sections and recombination coefficients were given
in \S\ref{simple}; the charge exchange rates were derived from the fits
by Kingdon \& Ferland \markcite{3407} (1996) to experimentally
determined cross sections \markcite{3420} (Marr \& West 1976).

For realistic cases in the ISM, the radiation is usually distributed
over a broad range of energy, and one must solve
Eqs.~\ref{n0}$-$\ref{n(e)} with appropriate values of $\Gamma$. 
Weighted averages of the numbers in Table~\ref{Pprimes} for the
respective contributions at various energies can give an approximate
indication for $P_{\rm Ar}^\prime$ in a complex radiation field.

\placetable{Pprimes}
\begin{deluxetable}{rccccc}
\tablewidth{0pt}
\tablecolumns{6}
\tablecaption{Parameters for the Relative Ionization of Argon to
Hydrogen\label{Pprimes}}
\tablehead{
\colhead{} & \colhead{} & \colhead{}&
\multicolumn{3}{c}{$\log P_{\rm Ar}^\prime$\tablenotemark{a}~~for
indicated}\\
\colhead{Created Ions} & \colhead{$E$ (eV)} & \colhead{$\log P_{\rm
Ar}$}& \multicolumn{3}{c}{values of $\Gamma({\rm H})/n(e) [{\rm
cm}^3{\rm s}^{-1}]$}\\
\cline{4-6}\\
\colhead{} & \colhead{} & \colhead{}& \colhead{$10^{-12}$} &
\colhead{$10^{-13}$} & \colhead{$10^{-14}$}\\}
\startdata
H$^+$&14&$-\infty$&$-\infty$&$-\infty$&$-\infty$\nl
H$^+$, Ar$^+$&16&0.764&0.761&0.732&0.518\nl
H$^+$, Ar$^+$&20&1.086&1.083&1.054&0.839\nl
H$^+$, Ar$^+$&24&1.310&1.306&1.278&1.063\nl
H$^+$, Ar$^+$, He$^+$&25&1.347&1.329&1.243&1.021\nl
H$^+$, Ar$^+$, Ar$^{++}$, He$^+$&28&1.410&1.407&1.301&1.072\nl
H$^+$, Ar$^+$, Ar$^{++}$, He$^+$&35&1.246&1.239&1.131&0.895\nl
H$^+$, Ar$^+$, Ar$^{++}$, He$^+$&40&0.925&0.913&0.812&0.581\nl
H$^+$, Ar$^+$, Ar$^{++}$, He$^+$&47&0.593&0.581&0.492&0.281\nl
H$^+$, Ar$^+$, Ar$^{++}$, He$^+$&54&0.813&0.799&0.697&0.462\nl
H$^+$, Ar$^+$, Ar$^{++}$, He$^+$,
He$^{++}$&55&0.855&0.831&0.712&0.493\nl
H$^+$, Ar$^+$, Ar$^{++}$, He$^+$,
He$^{++}$&70&1.313&1.292&1.149&0.900\nl
H$^+$, Ar$^+$, Ar$^{++}$, He$^+$,
He$^{++}$&100&1.728&1.719&1.551&1.283\nl
\enddata
\tablenotetext{a}{To be used in Eq.~\protect\ref{DeltaD} with a
substitution of $P_{\rm Ar}^\prime$ for $P_{\rm Ar}$.}
\end{deluxetable}
\clearpage

\subsubsection{Interpretation of the Deficiency of Ar~I toward
$\beta$~Cen~A}\label{beta_cen}

The largest deficiency of argon that we observed with IMAPS was for the
line of sight toward $\beta$~Cen~A.  If the true abundance of Ar in the
local ISM is about equal to the weighted average for B stars and the Sun
that we derived in \S\ref{abundances}, then the deficiency $D$ is equal
to the value shown in Table~\ref{abund}.  An alternative interpretation
is that the abundance of Ar in the ISM is depressed below stellar
abundances by an amount that could be as large as the value of $D$ that
we observed for $\gamma^2$~Vel.  In that case, the reduction of the
neutral argon abundance toward $\beta$~Cen~A caused by ionization is
equal to only $-$0.43\,dex.  In the discussion that follows, we will
adhere to this more conservative viewpoint.

In order to achieve $\Delta D_{\rm Ar}$ of $-$0.43, Eq.~\ref{DeltaD}
tells us that $n(e)/n({\rm H})$ must be about 0.23 if $P_{\rm
Ar}^\prime\approx 10$.  For $T=5000$\,K we require that $\Gamma({\rm
H})=\big[n(e)/n({\rm H})\big]^2\alpha({\rm H})n({\rm H})$ must be
$2.4\times 10^{-15}$ or $2.4\times 10^{-14}\,{\rm s}^{-1}$ for an
assumed $n({\rm H})=0.1$ or $1.0\,{\rm cm}^{-3}$, respectively.  For
either value of $n({\rm H})$, $\Gamma({\rm H})/n(e)=1.0\times
10^{-13}\,{\rm cm}^3{\rm s}^{-1}$.  If all the material in front of
$\beta$~Cen~A were inside a monolithic cloud with no internal sources of
ionizing radiation, the opacity due to any substantial fraction of the
hydrogen column density $N({\rm H})=10^{19.54}\,{\rm cm}^{-2}$ would
make it difficult to achieve the required level of $\Gamma({\rm H})$
from any reasonable external source of ionizing photons.  We are
therefore forced to conclude that the gas in front of $\beta$~Cen~A is
broken into well separated fragments that each has a much lower column
density for the radiation to penetrate.  Just from the fact that we see
two separate velocity components in the argon absorption lines (see
Figs.~\ref{intens_prof} and \ref{na_prof}), we can conclude that the
number of clouds is at least two.  Another possibility, one that invokes
a more specially contrived geometry, is that our line of sight skims
through a long edge of a cloud, very close to a surface that is
illuminated from the side.

If we were to propose that the gas in front of $\beta$~Cen~A is
comprised of about 5 separate clouds, each bathed in an ionizing field
on all sides, the average atom will be shielded by approximately each
cloud's column density $N({\rm H})=(10^{19.54}\,{\rm cm}^{-2})/5$
divided by four, i.e., $N({\rm H})\approx 10^{18.2}\,{\rm cm}^{-2}$. 
For an incident EUV spectral energy distribution that is similar to a
Planck distribution with $T=17,000$\,K [e.g., the radiation from a star
like $\epsilon$~CMa -- see Vallerga \& Welsh \markcite{298} (1995), or
perhaps $\beta$~Cen~A itself], the average energy of the photons that
penetrate through the hydrogen is $\langle E\rangle=20$\,eV.  The
strength of this radiation would need to be equivalent to the flux from
one such star at an unreasonably small distance of about 30\,pc to
achieve the required $\Gamma({\rm H})$ for $n({\rm H})=0.1\,{\rm
cm}^{-3}$ (or 9\,pc for $n({\rm H})=1.0\,{\rm cm}^{-3}$), but this
requirement will be eased if there is a significant additional
contribution from white dwarf stars or the soft x-ray radiation produced
by hot gases in the intercloud medium.  From Table~\ref{Pprimes} we
learn that $\log P_{\rm Ar}^\prime=1.054$ when $\langle E\rangle=20$\,eV
and $\Gamma({\rm H})/n(e)=10^{-13}\,{\rm cm}^3{\rm s}^{-1}$, and this
value of $P_{\rm Ar}^\prime$ is consistent with our original assumption
that led to  $\Delta D_{\rm Ar}=-0.43$ when $n(e)/n({\rm H})=0.23$. 
Additional sources of much harder radiation will raise $\langle
E\rangle$ which may in turn increase $P_{\rm Ar}^\prime$ and make it
easier to achieve our measured $\Delta D_{\rm Ar}$.

\subsubsection{Ar~I as a Discriminant for the Source of Ionization in
the Local Interstellar Cloud}\label{lism}

We digress to consider the potential importance of Ar~I abundances for
our understanding of the very local, partially ionized interstellar
medium that surrounds our solar system, just outside the heliosphere. 
The photoionization of hydrogen and helium in the Local Interstellar
Cloud (LIC) is thought to be strongly dominated by radiation from
$\epsilon$~CMa \markcite{298} (Vallerga \& Welsh 1995), nearby white
dwarf stars  (Dupuis et al. 1995), and possibly line emission from hot
($T\approx 10^6$\,K) gas that surrounds the LIC \markcite{1834} (Cheng
\& Bruhweiler 1990) or radiation from the conductive interface at the
boundary between the LIC and this hot gas \markcite{1773} (Slavin 1989). 
As we summarize below, the simple picture of an equilibrium established
by photoionization has problems that are difficult to reconcile with
observations.  Thus, it is worthwhile to investigate whether or not
future measurements of $N({\rm Ar~I})$ to nearby stars can help us to
understand better the real nature of the ionizing processes.

Vallerga \markcite{3481} (1998) has combined {\it EUVE\/} measurements
of 54 of the brightest stars that emit ionizing radiation in our
vicinity to arrive at photoionization rates $\Gamma({\rm He})=
8.25\times 10^{-16}\,{\rm s}^{-1}$ and $\Gamma({\rm H})= 1.56\times
10^{-15}\,{\rm s}^{-1}$.  The ratio of the former to the latter is not
sufficient to explain the observations that indicate that the fractional
ionization of helium seems to be as large as, or slightly greater than,
that of hydrogen \markcite{3171} (Dupuis et al. 1995), especially since
$\alpha({\rm He})>\alpha({\rm H})$ over all temperatures of interest. 
Vallerga \markcite{3481} (1998) estimates that the radiation field
arising from a large number of unobserved late-type stars could raise
$\Gamma({\rm He})$ by approximately 14\%, but clearly this increase is
too small to explain the large ionization of He. 

A reasonable proposal for overcoming the He ionization problem is one
stated by Cheng \& Bruhweiler \markcite{1834} (1990), where EUV
radiation from hot gases known to be emitting soft x-rays provides the
needed photons that are more energetic than the He ionization potential.
Unfortunately, this interpretation is upset by the inability of Jelinsky
et al. \markcite{3442} (1995) to observe this diffuse radiation during
long, dark-sky integrations with EUVE.  Jelinsky et al. concluded that
their upper limits for the energetic line and continuum emission, if
expanded to include the whole sky, translated to $\Gamma({\rm He})<
1.4\times 10^{-15}\,{\rm s}^{-1}$, an upper limit that is well below a
value that would be sufficient to explain the ionization of He. 
Possible reasons for this discrepancy between the observations and
simple theoretical predictions for the emission by a hot plasma may be
attributable to non-equilibrium effects or a depletion of some heavy
elements that are key contributors to the EUV line emission.  Lyu \&
Bruhweiler \markcite{3727} (1996) proposed a possible solution to the
helium ionization problem: the LIC was ionized at some time in the past,
possibly by the passage of a shock from a supernova, and the atoms
within the LIC have not yet reached their equilibrium states of lower
ionization.  The time constant for such an equilibrium to be reached
after the ionization is turned off is $\alpha({\rm He})n(e)\approx
1.8\times 10^{-14}\,{\rm s}^{-1} = 1/(1.8\times 10^6\,{\rm yr})$.

To demonstrate how the abundance of Ar~I could give further insights on
this ionization problem, we calculated photoionization equilibria using
the methods outlined in \S\ref{complex}, assuming that the radiation
field of Cheng \& Bruhweiler \markcite{1834} (1990) is in some sense
correct but that, somehow, we are misinterpreting the restrictions on
the steady-state flux imposed by the observations by Jelinsky et al
\markcite{3442} (1995).  In this exercise, we took the local ionization
rates arising from the local stellar radiation field calculated by
Vallerga \markcite{3481} (1998) and added them to those of an
approximation\footnote{For the EUV hot plasma emission lines, we
substituted $\delta$-functions at 35~eV and 69~eV and made them match
the specified $\Gamma({\rm He})$ and $\Gamma({\rm He}^+)$. To this we
added a flat spectrum (vs. $E$) between the H and He ionization edges to
simulate the contribution from thermal bremstrahlung.} to the radiation
field from the hot gas calculated by Cheng \& Bruhweiler \markcite{1834}
(1990) after it has been absorbed by $N({\rm H})=10^{18.30}\,{\rm
cm}^{-2}$ plus $N({\rm He})=10^{17.05}\,{\rm cm}^{-2}$, giving the
photoionization rates shown in Table~\ref{condx}.  From
Eqs.~\ref{n0}$-$\ref{n+} with $X$=He, we found equilibrium conditions
for the local gas that are shown in the top portion of
Table~\ref{condx}.  The effective path length of neutral gas that
absorbs the ionizing radiation from the hot gas and the local electron
density were both adjusted to give the best agreement with observed
values of $n({\rm H})=0.16\,{\rm cm}^{-3}$ \markcite{2741, 2780, 3301}
(Ajello et al. 1994; Qu\'emerais et al. 1994; Qu\'emerais, Sandel, \& de
Toma 1996), $f_+({\rm He})=0.25$ \markcite{2476} (Vennes et al. 1993)
and $n({\rm He})/n({\rm H})=0.07$ \markcite{3171} (Dupuis et al. 1995). 
However our value of $n(e)=0.036\,{\rm cm}^{-3}$ that is consistent with
these constraints turns out to be somewhat lower than the determination
of $0.11_{-0.06}^{+0.12}\,{\rm cm}^{-3}$ by Wood \& Linsky
\markcite{3430} (1997) and considerably lower than determinations from
various measurements of Mg ion fractions \markcite{2730, 3097, 3440}
(Frisch 1994; Gry et al. 1995; Lallement \& Ferlet 1997).  It is
possible that these ionization measurements are strongly influenced by
material closer to the edge of the local cloud that is more ionized
because there is has less shielding of the radiation \markcite{3481}
(Vallerga 1998). 

\placetable{condx}

When we use Eqs.~\ref{n0}$-$\ref{n+} to solve for the argon ionization
under conditions summarized in Table~\ref{condx}, we find that $f_0({\rm
Ar})=0.37$, and $f_+({\rm Ar})=0.63$ (with virtually no ${\rm
Ar}^{++}$).  Substituting the values of $f_0({\rm Ar})$, $n(e)$, $n({\rm
H}^+)$ and $n({\rm H})$ into Eq.~\ref{Pprime} gives $P_{\rm
Ar}^\prime=8.0$.  Eq.~\ref{DeltaD} with $P_{\rm Ar}^\prime$ in place of
$P_{\rm Ar}$ yields $\Delta D_{\rm Ar}=-0.355$.  It will be interesting
to see if this prediction is confirmed by future observations of the
Ar~I absorption lines in the spectra of very nearby stars by IMAPS or
the Far Ultraviolet Spectroscopic Explorer (FUSE, to be launched in late
1998).  If it is, one may find it difficult to abandon the notion that
the material is close to ionization equilibrium.  If $\Delta D_{\rm
Ar}\approx 0$, then one could suppose that either (1) the gas was mostly
ionized by a burst of photons at some time in the past, leaving the H
and He to recombine at approximately equal rates, or (2) the gas was
once collisionally ionized and has now cooled somewhat, but it has not
yet had a chance to recombine and reach a new equilibrium (see
\S\ref{summary} below).
\begin{deluxetable}{rl}
\tablewidth{0pt}
\tablecaption{Conditions in the Local Cloud\label{condx}}
\tablehead{
\colhead{Property} & \colhead{Quantity}\\}
\startdata
$T$&7000\,K\nl
$n(e)$&0.036\,${\rm cm}^{-3}$\nl
$n({\rm H})$&0.16\,${\rm cm}^{-3}$\nl
$n({\rm H}^+)$&0.029\,${\rm cm}^{-3}$\nl
$n({\rm He})$&0.013\,${\rm cm}^{-3}$\nl
$n({\rm He}^+)$&0.0060\,${\rm cm}^{-3}$\nl
$n({\rm He}^{++})$&0.00026\,${\rm cm}^{-3}$\nl
$\Gamma({\rm H})$&$2.3\times 10^{-15}\,{\rm s}^{-1}$\nl
$\Gamma({\rm He})$&$9.7\times 10^{-15}\,{\rm s}^{-1}$\nl
$\Gamma({\rm He}^+)$&$4.5\times 10^{-15}\,{\rm s}^{-1}$\nl
$\Gamma({\rm Ar})$&$3.1\times 10^{-14}\,{\rm s}^{-1}$\nl
$\Gamma({\rm Ar}^+)$&$2.4\times 10^{-14}\,{\rm s}^{-1}$\nl
\enddata
\end{deluxetable}
\clearpage
\placetable{other_elem}
\begin{deluxetable}{ccccc}
\tablecaption{Ionization Parameters for Mostly Neutral Elements in the
Local Cloud\label{other_elem}}
\tablehead{
\colhead{Element} & \colhead{$\log f_0$\tablenotemark{a}} &
\colhead{$\log P$\tablenotemark{b}} & \colhead{$\log
P^\prime$\tablenotemark{c}} & \colhead{$\Delta D$\tablenotemark{d}}\\}
\startdata
He&$-$0.173&0.441&0.387&$-$0.101\tablenotemark{e}\nl
N&$-$0.122&0.906&0.221&$-$0.049\nl
O&$-$0.076&1.123&0.019&$-$0.004\nl
Ne&$-$0.674&1.383&1.244&$-$0.602\nl
Ar&$-$0.427&1.013&0.902&$-$0.355\nl
\enddata
\tablenotetext{a}{See Eq.~\protect\ref{n0}.}
\tablenotetext{b}{See Eq.~\protect\ref{Peq}.}
\tablenotetext{c}{See Eq.~\protect\ref{Pprime}.}
\tablenotetext{d}{See Eq.~\protect\ref{DeltaD}, but substitute
$P^\prime$ for $P$.}
\tablenotetext{e}{$n(e)$ and the overall strength of radiation from the
hot gas were adjusted to make $\Delta D_{\rm He}$ agree approximately
with the constraints $n({\rm H})=0.16\,{rm cm}^{-3}$, $f_+({\rm
He})=0.25$, and $n({\rm He})/n({\rm H})=0.07$ in the local cloud -- see
text for details.}
\end{deluxetable}
\clearpage

We have performed similar calculations for other abundant elements that
are expected to be primarily neutral: He, N, O and Ne.  The results of
those calculations (plus a repeat of the information pertaining to He
and Ar) are given in Table~\ref{other_elem}.  We draw attention to the
fact that for N and O the values for $P^\prime$ are close to 1 and thus
$\Delta D\approx 0$, making these two elements good substitutes for H if
$N({\rm H})$ is not easy to measure.  This result is a consequence of
the very large charge transfer rates for these elements with
hydrogen\footnote{The very close match of the ionization potentials of H
and O leads to the very strong charge exchange reactions that go easily
in both directions, so that $f_0({\rm O})\approx f_0({\rm H})$
\protect\markcite{3406} (Field \& Steigman 1971).  For N, only the
reaction ${\rm N}^++{\rm H}\rightarrow {\rm N} + {\rm H}^+$ goes easily
and it is not as strong as that for O \protect\markcite{1927} (Butler \&
Dalgarno 1979), so if $n(e)\gg n({\rm H})$, $P_{\rm N}^\prime\rightarrow
P_{\rm N}$.}, essentially locking their ionization ratios to nearly that
of hydrogen.  It is clear that the simple $P$ values for these elements
can be very misleading.

Measurements of pick-up ions \markcite{2648} (Geiss et al. 1994) and
anomalous cosmic rays\footnote{Pick-up ions and anomalous cosmic rays
(ACRs) are believed to originate primarily from the neutral component of
the interstellar gas that can penetrate well inside the heliosphere
(charged atoms are deflected).  Once inside, the atoms are ionized by
either charge exchange with solar wind protons or UV ionizing radiation
from the Sun.  They are then swept up by the solar wind (hence the name
``pick-up ions'') and deposited in its termination shock, where they are
subsequently accelerated to energies of about 5$-$50\,MeV~nucleon$^{-1}$
to become anomalous cosmic rays.} \markcite{3445} (Fisk, Kozlovsky, \&
Ramaty 1974) have already given us some indications about the relative
abundances of neutral species in the local cloud.  The pick-up ion
ratios He/O, N/O and Ne/O measured by Geiss, et al. \markcite{2648}
(1994) seem to be consistent, within the experimental errors, with the
respective cosmic abundance ratios.  Similarly, Cummings \& Stone
\markcite{3444} (1995) found that the abundances of the He, N, O, Ne and
Ar components of anomalous cosmic rays seemed to reflect their cosmic
abundances.  To obtain these results, however, Cummings \& Stone
\markcite{3444} (1995) had to incorporate an empirical correction for
the efficiency of the acceleration and propagation processes in the
heliosphere.  This correction was simply a power-law in atomic mass,
adjusted to make the abundances of He and Ne agree with their respective
cosmic ratios.  Even with this adjustable parameter, it appears that the
values predicted for photoionization equilibrium given in
Table~\ref{other_elem} are not easy to reconcile with the observations
of anomalous cosmic rays, i.e., He, Ne and Ar do not seem to be
depressed below their cosmic abundances relative to O and N.  It is
possible that a modification of O and N abundances by the strong charge
exchange reactions with solar wind protons in the heliospheric interface
\markcite{2897} (Fahr, Osterbart, \& Rucinski 1995) may effectively
cancel the relative reductions in the noble gas abundances predicted in
Table~\ref{other_elem}.

\section{Summary}\label{summary}

We have measured the abundances of neutral argon relative to neutral
hydrogen in the ISM over 3 different lines of sight (and obtained lower
limits for 5 others).  These abundances are lower than those measured in
B-type stars or the Sun, repeating an effect seen for other elements
that do not have strong depletions caused by dust formation.  In
contrast to the relatively constant ISM abundances of these other
lightly depleted elements, our determinations indicate that argon's
relative abundance seems to vary from one place to the next.

We have stated theoretical arguments to suggest that, for our lines of
sight, it is unlikely that an appreciable amount of argon is bound into
the dust grains that accompany the gas.  Instead, the missing Ar can be
explained by most of the atoms being ionized and thus becoming invisible
to us in this survey.  The very large photoionization cross section of
neutral argon, relative to that of hydrogen, can result in argon being
ionized more thoroughly than hydrogen in partially ionized regions.  For
a line of sight that shows the largest reduction of Ar (that toward
$\beta$~Cen~A) we conclude that the cloud must be broken into at least
about 5 subclouds to allow radiation from the outside to penetrate most
of the material and cause the effect that we observe.
 
Krypton is the only other noble gas that has been well studied in the 
neutral ISM \markcite{2682, 3424} (Cardelli 1994; Cardelli \& Meyer
1997), and it too has a large cross section for ionization:
$\sigma\approx 3\times 10^{-17}\,{\rm cm}^{-2}$ above the ionization
threshold \markcite{3420} (Marr \& West 1976).  Because Kr has such a
low cosmic abundance [$\log ({\rm Kr/H})=-8.77$, \markcite{68} (Anders
\& Grevesse 1989)] the UV absorption feature is very weak.  Thus, the
studies of Kr/H have featured sight lines that have $\log N({\rm
H~I})\geq 20.5$, and regions with this much column density are expected
to be fully shielded and neutral, except for those elements with
ionization potentials below that of H.  Hence it is no surprise that the
Kr investigations do not uncover the abundance fluctuations that we see
here for Ar.

The ratio of Ar~I to H~I (or some other element that is not usually
depleted in the ISM) may offer a critical insight on the nature of the
ionizing process that is responsible for the many neutral clouds that
seem to be unexpectedly rich in free electrons.  If such clouds are
ionized mainly by photons, we should expect to see a reduction in the
apparent abundance of Ar~I.  Alternatively, if the Ar~I seems normal,
partial recombination from a fully ionized condition or the effects of
collisional ionization may be more plausible interpretations
\markcite{356} (Trapero et al. 1996).  For instance, in collisional
ionization equilibrium up to temperatures of around $10^{4.4}$\,K, the
fractional abundances of argon in its neutral form are of order or
slightly more than those of oxygen \markcite{1936} (Shull \& Van
Steenberg 1982).  Above $10^{4.4}$\,K, $f_0({\rm Ar}) < f_0({\rm O})$,
but both elements have less than 1\% of their atoms in neutral form at
these temperatures.  For a gas that is cooling either isochorically or
isobarically with some lag in the recombination, the fractional
ionizations of O and Ar are nearly identical for $f_0>10^{-3}$ (which
means that $T<10^{4.5}$\,K) \markcite{327} (Benjamin \& Shapiro 1997).

In future studies of UV absorption lines, it will be interesting to
confirm the validity of our proposal that partially ionized regions
offer an explanation for the reduction in Ar~I.  For instance, is there
a correlation between $D({\rm Ar})$ and $n(e)$ (as indicated by the
fine-structure excitations of C~II or Si~II) or the existence of atoms
in moderately high stages of ionization, such as Si~III or S~III?  If
such correlations hold true, then we can use the Ar~I abundances as
either a powerful diagnostic on the nature of the ionization or,
alternatively, as an indicator of whether the higher ions arise from a
completely ionized classical H~II region (where there should be little
Ar~I or H~I) or a partially ionized cloud of the type described here.

\acknowledgments

Support for flying IMAPS on the ORFEUS-SPAS-I mission and the research
carried out by EBJ came from NASA Grant NAG5-616 to Princeton
University.  UJS performed most of his work on this project while he
held a National Research Council - NASA/GSFC Research Associateship, and
additional support was provided by the NASA-LTSARP grant NAG5-3539 to
Villanova University.  Most of the development of the data reduction
software was done by EBJ shortly after the mission, while he was
supported by a research award for senior U.S. scientists from the
Alexander von Humboldt Foundation and was a guest at the Institut f\"ur
Astronomie und Astrophysik in T\"ubingen, Germany.  We thank Drs.
B.~T.~Draine, T.~M.~Tripp and J.~V.~Vallerga for their comments on an
early draft of this paper.

The ORFEUS-SPAS project was a joint undertaking of the US and German
space agencies, NASA and DARA.  The successful execution of our
observations was the product of efforts over many years by engineering
teams at Princeton University Observatory, Ball Aerospace Systems Group
(the industrial subcontractor for the IMAPS instrument) and Daimler-Benz
Aerospace (the German firm that built the ASTRO-SPAS spacecraft and
conducted mission operations).

\end{document}